\documentclass[showpacs,superscriptaddress,preprint]{revtex4-1}
\usepackage{amsmath}
\usepackage{bm}
\pdfoutput=1
\usepackage{epsfig}
\usepackage{epstopdf}
\usepackage{natbib}
\usepackage[utf8]{inputenc}
\usepackage{gensymb}

\newcommand{\tref}[1]{Table~\ref{#1}}

\newcommand{\fermi}{G_\mathrm{F}}

\newcommand{\podd}{$\cal P$-odd~}
\newcommand{\ptodd}{$\cal P,T$-odd~}
\newcommand{\sdpv}{$\cal NSD-PV$~}

\onecolumngrid

\begin{document}

\title{Laser-coolable polyatomic molecules with heavy nuclei}
\author{T.A.\ Isaev}
\affiliation{Petersburg Nuclear Physics Institute,
             Orlova Roscha. 1, 188300 Gatchina, Russia}
\author{A.V.\ Zaitsevskii}
\affiliation{Petersburg Nuclear Physics Institute,
             Orlova Roscha. 1, 188300 Gatchina, Russia}
\affiliation{Chemistry Dept., M. Lomonosov Moscow State University, Moscow 119991, Russia}
\author{E.\ Eliav}
\affiliation{School of Chemistry, Tel-Aviv University, 66978, Tel-Aviv, Israel}
\date{\today}

\begin{abstract}
Recently a number of diatomic and polyatomics molecules has been identified
as a prospective systems for Doppler/Sisyphus cooling.
Doppler/Sisyphus cooling allows to decrease the kinetic energy of molecules down
to microkelvin temperatures with high efficiency
and then capture them to molecular traps, including magneto-optical trap. Trapped molecules can be used for
creation of molecular fountains and/or
performing controlled chemical reactions, high-precision spectra measurements  and a multitude of other applications.
Polyatomic molecules with heavy nuclei present considerable interest for the search for ``new physics'' outside of
Standard Model and other applications including cold chemistry, photochemistry,
quantum informatics etc. Herein we would like to attract attention to radium monohydroxide molecule
(RaOH) which is on the one hand an amenable object
for laser cooling and on the other hand provides extensive
possibilities for searching for \podd and \ptodd effects. At the moment RaOH is the heaviest polyatomic molecule proposed for
direct cooling with lasers.
\end{abstract}

\maketitle

\section*{Introduction}
The scheme of molecular Doppler cooling (first displayed in \cite{DiRosa:04}) proved to be
efficient, robust and applicable to a large class of molecular species.
Recently highly-efficient cooling and magneto-optical trapping of diatomic molecules
was demonstrated for SrF \cite{Barry:2014}, YO \cite{Hummon:2013} and CaF \cite{Hemmerling:2016,
Anderegg:2017}.
The experiments on Doppler/Sisyphus cooling
on BaH \cite{Tarallo:2016}, MgF \cite{Liang:2016} and some other molecules and their cations
are currently conducted by different groups. Recently we have proposed a number of
polyatomic molecules containing light atoms, which may be expected
to be suitable for Doppler/Sisyphus cooling \cite{Isaev:2016}. The estimate for Doppler limit temperature for considered
molecules (CaOH, CaNC, MgCH$_3$, CaCH$_3$ and chiral MgCHDT) showed that it is quite
realistic to reach the submillikelvin temperatures with a large class of polyatomic species
(see also recent proposal by I. Kozyryev et al \cite{Kozyryev:2016b} for
a large class of polyatomic radicals amenable for laser cooling).
In 2017 J. Doyle's team at Harvard University has reported the first observation
of fast Sisyphus laser cooling of SrOH to the temperature of $\sim$750 $\mu K$ \cite{Kozyryev:2017a}.
To our knowledge the only competitive method currently allowing
to cool polyatomic (four- and bigger multi-atomic
molecules) to submillikelvin temperatures is
optoelectrical Sisyphus cooling method  \cite{Prehn:2016, Zeppenfeld:2012}. The latter method,
though having great potential,
relies on strong Stark interaction with external field and is expected to be most effective
when applied to molecules with rather large dipole moments. The method of molecular laser cooling
can be applied both to dipolar and non-dipolar species and, besides, is directly
related to very well-studied,
efficient and robust atomic Doppler cooling technique. On the one hand
electronic structure of heavy-atom molecules especially favors quasi-diagonal structure
of Franck-Condon (FC) matrix between ground and excited electronic states \cite{isaev:2010}
due to more diffuse valence electronic orbital in comparison with light-atom homologies.
On the other hand the molecular species containing
heavy atoms present considerable interest for the experiments devoted to
search for new physics outside the Standard
Model \cite{Titov:06amin, Klemperer:1993,Kozyryev:2017b},
including search for axion-like-particle candidates for dark matter
\cite{Stadnik:2014} and search for parity violation in chiral molecules (see e.g \cite{Berger:1997}).
Herein we apply the principles for identifying the laser-coolable polyatomic species
introduced in \cite{Isaev:2016} to the field of heavy-atom compounds and
propose the triatomic molecule RaOH
as a candidate for Doppler cooling and for search for
space parity violating (\podd) and
also for both space parity and time-reversal violating (\ptodd) interactions
with polyatomic molecules. In the first part of the article
 we calculate molecular spectroscopic parameters for
this molecule and estimate the completeness
of the cooling loop by calculating the FC matrix for vibronic transitions
between first electronic excited and ground states.
Then we apply recently developed original methodology for
calculations of electric transition dipole moments between
elaborate electronic levels in fully-relativistic framework.
Finally, we use the earlier developed by us approach to
estimate \podd and \ptodd parameters of effective molecular
spin hamiltonian in the framework of 
Zero Order Regular Approximation (ZORA) \cite{Isaev:2012, Isaev:2014} 
(see also \cite{berger:2005, berger:2005a}).

\section*{Relativistic electronic structure calculations}

\subsection*{Equilibrium molecular parameters}
As it was pointed out in \cite{isaev:2010} the crucial prerequisite for the closed cooling transition
loop in molecules (required for the quasi-diagonality
of the FC matrix) is the conservation of a similar molecular geometry for both ground and first excited electronic states. This happens for e.g.
transition of the valence electron between non-bonding orbitals being presented
in the leading electronic configuration for both ground end excited electronic states.
In \cite{Isaev:2016} we have proposed a
simple scheme for obtaining the molecular structures of polyatomic species which contain
non-bonding single-occupied molecular orbitals. The basic idea is to make a substitution of the halogen in the series of
MX molecules (M is the metal and X is the halogen) to pseudo-halogen
or functional groups. In our consideration in \cite{Isaev:2016} the heaviest nucleus in a molecule was calcium
(with the charge number $Z$=20) and, thus we
were able to safely neglect the relativistic effects in those
electronic structure calculations. For light
element compounds such an approximation indeed does not lead to noticeable
changes in molecular parameters (see e.g. \cite{DyallRelBook}).

\begin{figure*}[h]
\label{pes}
\caption{ (Color online)
Potential energies of the ground and first excited electronic state as functions of $R_{\rm Ra-O}$ (left) and the valence Ra-O-H angle (rigrh) with other parameter fixed at their equilibrium values
(see supplementary material for the details). Energies are given with respect to the ground state
equilibrium point. Solid black lines: DC/FS-RCCSD, dashed blue lines:  SF/FS-RCCSD.
}
\includegraphics[width=0.75\linewidth]{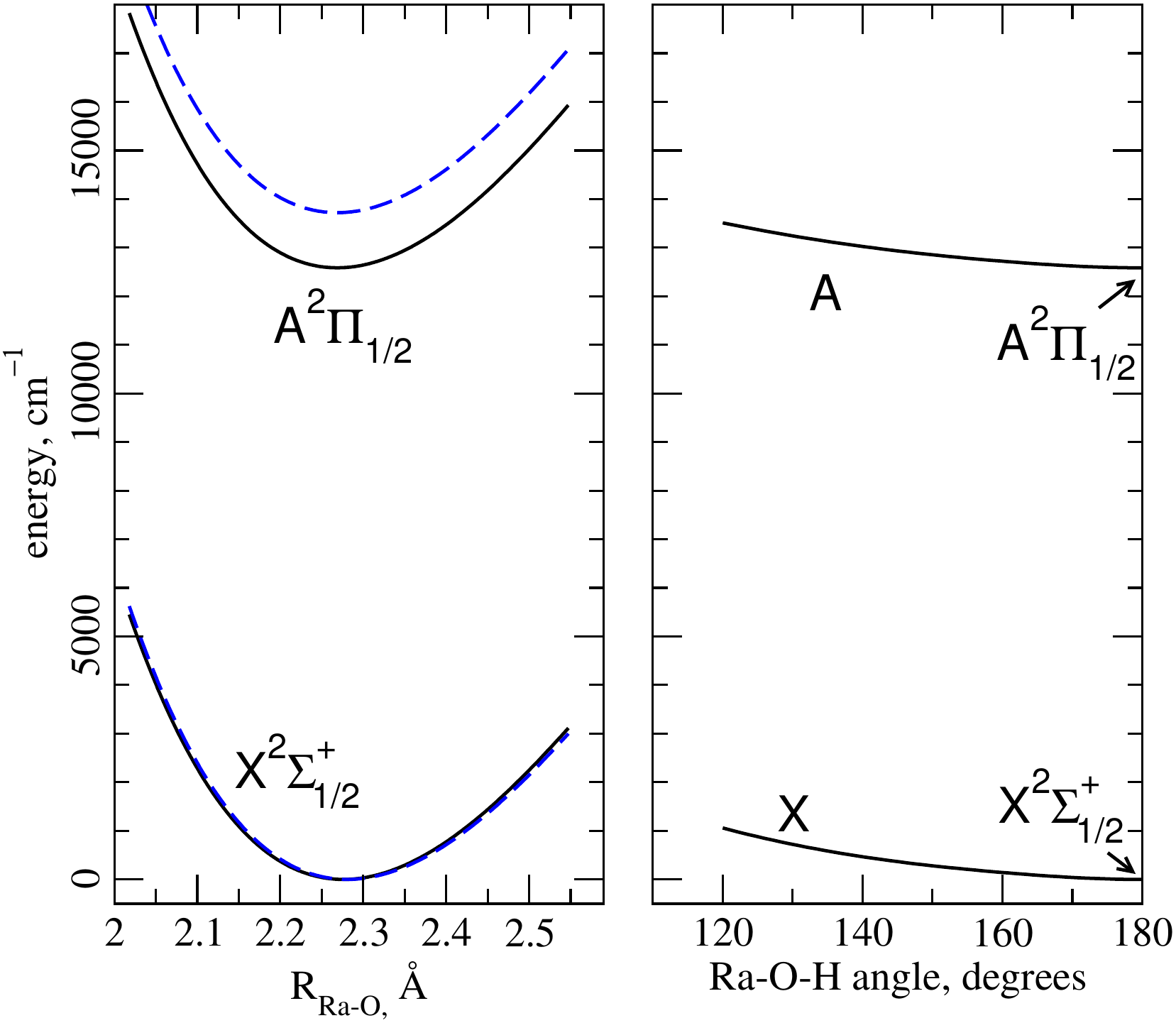}
\end{figure*}

In the opposite case of heavy element
compounds one has to estimate the influence of
relativistic effects on FC factors. FC factors are known to be strongly dependent on
displacements in equilibrium geometries and thus are
sensitive to both scalar and spin-dependent relativistic effects,
if such effects are
different for the considered electronic states (in our case the ground and
first excited electronic states).
We started our study from calculation of equilibrium structures and normal modes for both
ground and first excited electronic states of RaOH using spin-averaged relativistic effective
core potential ({\sc arecp}) with 10 valence electrons
treated explicitly for Ra and with all-electron description for O and H.
Multiconfiguration Self-Consistent Field method (MCSCF) is used for ARECP calculations with
17 valence electrons correlated and with
the basis sets of triple-zeta quality from Molpro program package
 library (see supplementary material for details).
The equilibrium structures and normal mode frequencies are provided in \tref{scal-fourcomp}.
It can be seen that RaOH
has essentially linear geometry in both ground ($^2\Sigma$, according to irreps of $C_{\infty v}$ group)
and first excited ($^2\Pi$) electronic states.
Very weak Renner -Teller effect in $^2\Pi$
RaOH detected within the spin-orbit-free approximation is in accord with similar
effects found
in light homologue compound CaOH \cite{Theodorakopoulos:1999,Li:1996} and should be fully suppressed due
to strong spin-orbit interaction in heavy-atom compounds.
We also would like to emphasize that in the vibrational analysis
we did not restrict symmetry of vibrational modes to the
totally-symmetric one.
To estimate the influence of the relativistic spin-orbit interaction on {\it relative displacements} in
equilibrium geometries we computed the properties for
the states of interest with scalar-relativistic four-component spin-free (SF) \cite{Dyall:1994} and
fully-relativistic four-component Dirac-Coulomb (DC) Hamiltonians.
In four-component calculations
we used Fock-Space Relativistic Coupled Cluster with Single and Double 
excitations (FS-RCCSD) method  to account for electron correlations \cite{Visscher:01}.
In this approach augmented
ANO-RCC basis set on Ra and double zeta plus polarization quality
basis sets by Alrichs et al on O and H (see supplementary material)
were used.
Thus we could also directly check the stability
of the calculated FC-factors in relation to changing of the method for electron correlation accounting.
 It follows from ARECP/MCSCF calculations that the main interest
present the vibrations localized on the Ra-OH bond.
Thus one has to consider first the cross-section of the potential energy surface
along Ra-OH bond
(while relative positions of O and H nuclei can be fixed). The bond angle Ra-O-H is expected
to be very close to $180^{\circ}$ due to above-mentioned suppression of Renner-Teller effect
by spin-orbit interaction. Nevertheless to additionaly confirm this
we have also explicitly calculated dependence of the
energies of the ground and first excited electronic states on Ra-O-H bond angle (see Fig \ref{pes})
in the framework of DC/RCCSD approach.
One can see from \tref{scal-fourcomp}, that the changes in {\it displacements} of
the equilibrium geometries for ground and excited electronic states of RaOH are rather minor,
while the method of accounting for electronic correlations is changed and
simultaneously the spin-orbit interaction is taken into account. To check directly the influence of dynamic
electronic correlations and the spin-orbit interaction on FC factors we have calculated these factors for $0' -0$, $0' -1$ and $0'-2$ transition 
of the quasimolecule RaX. Here $0'$, 0, 1 and 2 are the vibrational quantum numbers for the first excited electronic state (with prime)
and ground electronic state (without prime) correspondingly; mass of quasinucleus X is equal to the mass of
OH group (17 a.m.u.), while internuclear potential is equivalent to
that between Ra and O nuclei  from either ARECP/MCSCF or
SF/FS-RCCSD or DC/FS-RCCSD calculations (see supplementary material).
The calculations of FC factors were performed analogously to those from
\cite{isaev:2010} and the
results are presented in Table I in supplementary material.
Similarly to \cite{Isaev:2016} we are using here the sum of the three largest FC-factors in order to estimate the completeness of the cooling loop.
One can see from the Table I that for every calculation the sum of the three largest FC-factors
is greater than 0.99, thus it is reasonable to expect highly-closed cooling transition
$^2\Sigma_{1/2} \rightarrow ^2\Pi_{1/2}$  in RaOH.

\subsection*{Transition dipole moments}
Another prerequisite for the efficient molecular Doppler cooling is strongly allowed dipole transition between
working electronic states.
To estimate radiative lifetimes for the first excited $^2\Pi_{1/2}$ state on the FS-RCCSD level,
we evaluated the transition electric dipole moment (TDM) between this state and the ground
$^2\Sigma_{1/2}$ electronic state using the finite-filed method. The TDM
component values were derived from the finite-difference 
approximation
for the derivative matrix elements as
\begin{eqnarray}
{\rm TDM}_\eta&=&\!\left(E_{^2\Pi_{1/2}}-E_{^2\Sigma_{1/2}}\right)
\left<\tilde{\Psi}^{\perp\perp}_{ ^2\Pi_{1/2}} (F_\eta)\left|
\frac{\partial}{\partial F_\eta}\tilde{\Psi}_{^2\Sigma_{1/2}}(F_\eta)\right.\right>
\left|\begin{array}{l}\\_{F=0}\end{array}\right. \nonumber \\
&\approx&\!\left(E_{^2\Pi_{1/2}}-E_{^2\Sigma_{1/2}}\right)
\frac{
\left<\tilde{\Psi}^{\perp\perp}_{ ^2\Pi_{1/2}} (\Delta F_\eta)\left|
\tilde{\Psi}_{^2\Sigma_{1/2}}(-\Delta F_\eta)\right.\right>}{2\Delta F_\eta}
,\quad \eta=x,\,y,\,z
\end{eqnarray}
Here $F$ is the applied uniform electric field strength and $\tilde{\Psi}^{\perp\perp}$ and
$\tilde{\Psi}$ stand for left and right eigenvectors of the field-dependent
non-Hermitian FS-RCC
effective Hamiltonian acting in the field-free ($F=0$) model space.
The chosen numerical differentiation step size $\Delta F_\eta=5\cdot10^{-5}$ atomic units corresponded approximately to the center of the interval on the logarithmic scale in which the dependence of the resulting TDM values on the step size was negligible. 
Although the calculations involve only the effective Hamiltonian eigenvectors, i.\ e. the
model space projections of many-electron wavefunctions,
the resulting transition moment approximations implicitly
incorporate the bulk of the contributions from the remainder
(outer-space)
part of these wavefunctions \cite{Zaitsevskii:1998}.
The computational procedure has been implemented
within the {\sc DIRAC15} program package \cite{dirac:15}.
The lifetimes of several vibrational levels $v'$
 are estimated according to (see \cite{Matsushita:1987}):
\begin{equation}
\tau_{v'}=\frac{4.936\cdot 10^5}{|\mathrm{TDM}|^2\sum_{v}\mathrm{(FC)}_{vv'}\Delta\mathrm{E}_{vv'}^3},
\label{lifetime}
\end{equation}
where $\tau_{v'}$ is the radiative lifetime of the vibrational level $v'$ (in $s$); $|\mathrm{TDM}|^2$
is the sum of the squares of transition dipole moment moduli between one component of
the initial electronic state and both components of the final state
(in atomic units); $\mathrm{(FC)}_{vv'}$ is the FC-factor for the vibronic transition $vv'$; and
$\Delta\mathrm{E}$ is the energy interval between vibronic levels $vv'$ (in wavenumbers).
The TDM value is evaluated at the equilibrium geometry of the ground electronic state of RaOH (which is nearly coinciding with that of the excited state) from DC/FS-RCCSD calculations.
For $\tau_{0}$ we obtain the lifetime of the
excited electronic state about 40 $ns$ which corresponds to Doppler limit temperature
$T_\mathrm{D}\approx 90 \mu K$.

\begin{figure*}[h]
\caption{
\label{tdm}(Color online) Absolute value of $A^2\Pi_{1/2}-X^2\Sigma^+_{1/2}$ transition dipole moment $M$ as a function of $R_\mathrm{Ra-O}$ according to DC/FS-RCCSD calculations. Dashed line: longitudinal component, $|\langle A^2\Pi_{1/2,m}\vert M_z \vert X^2\Sigma^+_{1/2,m}\rangle |$ ($|$TDM$_{\parallel}|$); dotted line: transverse component $|\langle A^2\Pi_{1/2,1/2}\vert \sqrt{1/2}\,(M_x+{\rm i} M_y )\vert X^2\Sigma^+_{1/2,-1/2}\rangle |$ ($|$TDM$_{\perp}|$); dash-dotted line: total transition dipole length ($|$TDM$|$). The solid line corresponds to the spin-orbit-free $|$TDM$|$ limit from the SF/FS-RCCSD calculations.
}
\includegraphics[width=0.75\linewidth]{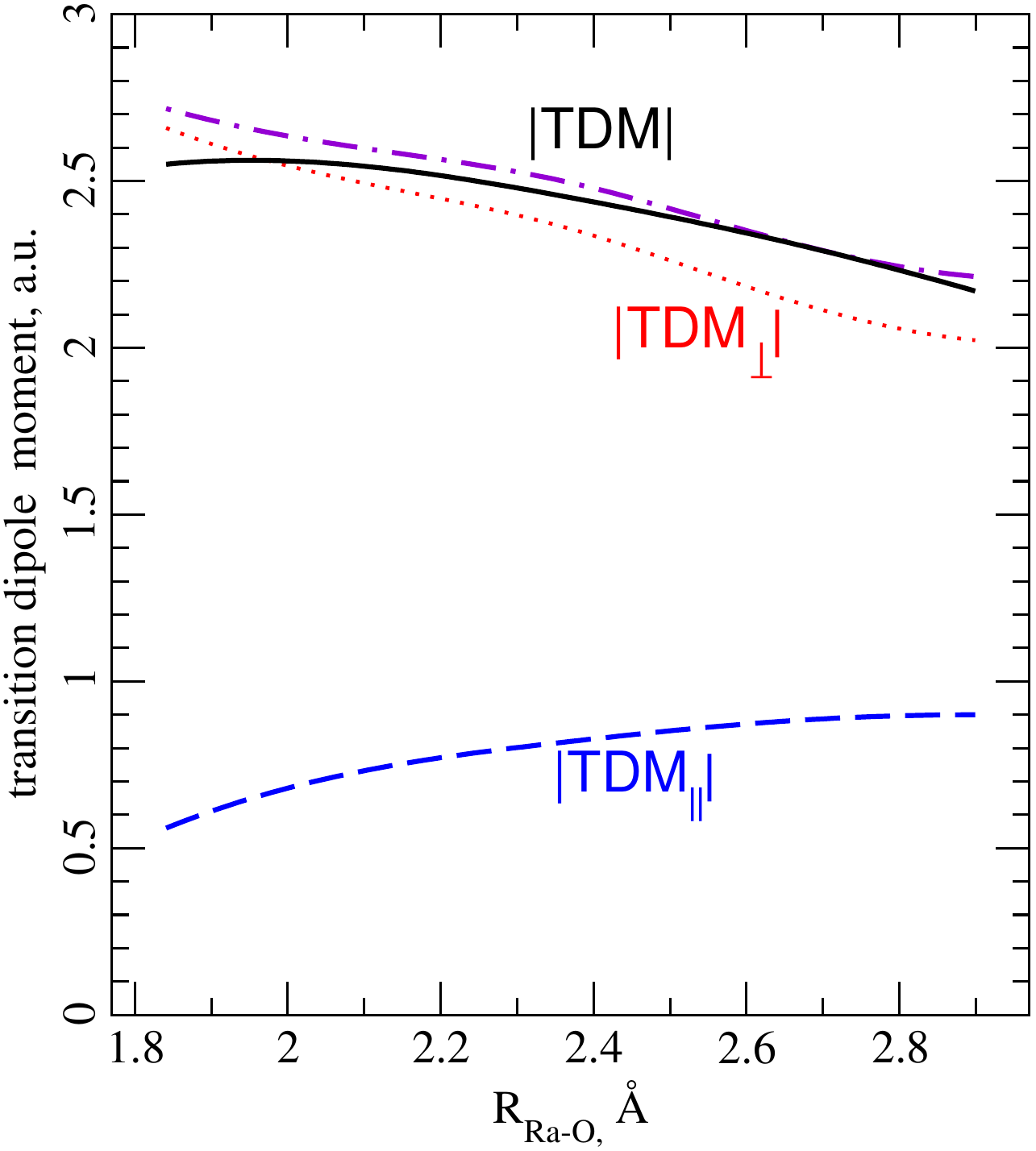}
\end{figure*}

 We have also calculated the dependence of the transition dipole moment (TDM) on
the internuclear distance Ra-O, to check the
behavior of TDM near the equilibrium Ra-O distance and
thus control possible Herzberg-Teller contribution to the vibrational spectrum
 (see Fig. \ref{tdm}). 
 This contribution is proportional to the derivation of the TDM
on the internuclear distance Ra-O.
To demonstrate the influence of the spin-dependent effects we
provide on the plot with TDMs from DC/FS-RCCSD calculations also the spin-orbit-free limit
of the total TDMfrom SF/FS-RCCSD calculations and
the longitudinal component  $\vert\mathrm{M}_z\vert$, which
vanishes at the scalar relativistic limit.
Rather slow and nearly linear variation of the
 TDM near the equilibrium geometries
of the ground and first excited electronic states in RaOH
justifies the use of the approximation (\ref{lifetime}) for radiative lifetimes.

\begin{table*}[th]
\caption{ \label{scal-fourcomp} Calculated molecular parameters and FC factors for RaOH molecule.
The results of ARECP/MCSCF, SF/FS-RCCSD and DC/FS-RCCSD calculations are provided.
Internuclear distances $R_i$ are given in \AA, electric dipole moments in Debye, transition wavenumbers $T_\mathrm{e}$ and harmonic vibrational
wavenumbers $\tilde{\omega}_\mathrm{e}$ of normal mode $\nu_l$ in inverse centimeters.
For FC factors the corresponding
vibrational quantum numbers are reported, e.g 1$^{0}_{2}$ is FC factor for a vibronic transition
between the ground (0) vibrational state of the excited electronic state and the vibrational state with the first mode (1)
(Ra-O stretching) being doubly excited (2) in the ground electronic state.
The irreducible representation $\mathrm{\pi}$ is two-dimensional, but the degeneracy could be slightly lifted in the numerical calculations, so we provide vibrational wavenumbers of both components.}
\begin{tabular}{lll|llll}
\vspace{0.2cm} \\
\multicolumn{7}{c}{ARECP/MCSCF} \\
State  &  \multicolumn{2}{c}{ } & &State  &  \multicolumn{2}{c}{ }  \\
\hline
X ($^2\Sigma$)      & $R_\mathrm{(Ra-O)} $     & 2.38  &  & A ($^2\Pi$)       &  $R_\mathrm{(Ra-O)}$    & 2.35     \\
                              & $R_\mathrm{(O-H)} $        & 0.94  &    &                       & $R_\mathrm{(O-H)}$    & 0.94           \\
                              & $\angle$ Ra-O-H               &  180.0 &   &                      &  $\angle$ Ra-O-H            & 180.0     \\
                              & $|D|$                                &  2.09  &      &                      & $|D|$                            &   1.70        \\
                             &                                         &             &     &                       & $T_\mathrm{e}$            &  12.0$\times$10$^{3}$     \\
                              \hline
\multicolumn{3}{c|}{Normal modes $\nu_l$}  & & \multicolumn{3}{c}{Normal modes $\nu_l$}   \\
X ($^2\Sigma$)     &  $\nu_3(\mathrm{\sigma}^+)$        &  4243  &    & A ($^2\Pi$)      & $\nu_3(\mathrm{\sigma}^+)$     &   4248    \\
                              &  $\nu_1(\mathrm{\sigma}^+)$        & 437     &      &              & $\nu_1(\mathrm{\sigma}^+)$          &  461      \\
                              &  $\nu_2(\mathrm{\pi})$                &  366/366  &      &           &  $\nu_2(\mathrm{\pi})$               & 383/383         \\
   \multicolumn{7}{c}{FC factors ~~ 0.9050\footnote{0$^{0}_{0}$}, 0.0922\footnote{1$^{0}_{1}$ Ra$-$OH stretching}, 0.0025\footnote{1$^{0}_{2}$ Ra$-$OH stretching}}         \\
     \multicolumn{7}{c}{$\sum >$ 0.99}     \\
                             \multicolumn{3}{c|}{{\sc sf/fs-rccsd}} & &  \multicolumn{3}{c}{{\sc dc/fs-rccsd}}  \\
 X ($^2\Sigma$)      & $R_\mathrm{(Ra-O)} $     & 2.30  &    & A ($^2\Sigma_{1/2}$)       &  $R_\mathrm{(Ra-O)}$    & 2.30    \\
 A ($^2\Pi$)             & $R_\mathrm{(Ra-O)} $     & 2.29  &    & A ($^2\Pi_{1/2}$)         &  $R_\mathrm{(Ra-O)}$    & 2.29    \\
                             &    $T_\mathrm{e}$              &    13.8$\times$10$^{3}$  &     &             & $T_\mathrm{e}$            &  12.6$\times$10$^{3}$     \\
\hline

\end{tabular}
\end{table*}%

\section*{Quasirelativistic calculations of P- and P,T-odd parameters}

We have previously derived the expressions for matrix elements of the \podd and \ptodd operators in
quazirelativistic Zero-Order Regular Approximation (ZORA) framework (see \cite{Isaev:2012} and
references therein) and related them to the corresponding
effective operators in polyatomic molecules with open shells.
Our previous calculations of \podd nuclear spin-dependent and \ptodd properties for diatomic molecules
within Generalized Hartree-Fock (GHF)
approach \cite{Isaev:2012} were later confirmed in
fully-relativistic four-component Dirac-Hartree-Fock (DHF) calculations \cite{borschevsky:2012, borschevsky:2013}.
It turned out that the difference between values of the \podd parameters in ZORA and four-component calculations
does not exceed 5\% (see \cite{borschevsky:2013}).
High-precision FS-RCC calculations of \podd and \ptodd molecular parameters for RaF
molecule \cite{kudashov:2014} are also in a good agreement with our previous estimates of influence of electronic correlation
on these parameters.

\begin{table*}[h]
\caption{\label{pvterms} \podd and scalar \ptodd term in ZORA Hamiltonian. The superscripts of $z$ denote
orders of the corrseponding ZORA terms on nuclear spin and \podd (\ptodd) coupling constants.}
 \begin{tabular*}{12cm}{lll}
 Term         & Name  & Expression \\
  \hline
  ~\\
   \multicolumn{3}{c}{\podd terms}  \\
 $z^{(0,1)}_\mathrm{s}$       & Scalar \podd interaction &
 $\displaystyle\frac{\fermi}{2\sqrt{2}} Q_A \lbrace\vec{\bm{\sigma}}\cdot \vec{p},
\frac{\tilde{\omega}}{c} \rho_A(\vec{r}) \rbrace$ \vspace{2mm} \\
 $z^{(1,1)}_\mathrm{hf}$      &    Scalar \podd  / hyperfine P-even interaction  &
 $\displaystyle\frac{\fermi}{2\sqrt{2}} Q_A \lbrace e\, \vec{\bm{\sigma}}\cdot\vec{A}_{\mu}(\vec{r}),\frac{\tilde{\omega}}{c} \rho_A (\vec{r})\rbrace$\vspace{2mm}\\
 $z^{(1,1)}_\mathrm{sd}$& Nuclear spin-dependent \podd interaction &
 $\displaystyle\frac{\fermi}{2\sqrt{2}} 2 k_{{\cal
A},A}\lbrace\vec{\bm{\sigma}}\cdot\vec{p}, \frac{\tilde{\omega}}{c}
 \vec{\bm{\sigma}}\cdot \vec{I}_A \rho_A (\vec{r}) \rbrace$ \vspace{2mm}\\
$z^{(2,1)}_\mathrm{sdr}$& Nuclear spin-dependent \podd / &
 $\displaystyle\frac{\fermi}{2\sqrt{2}} 2 k_{{\cal A},A}
\lbrace e\, \vec{\bm{\sigma}}\cdot\vec{A}_{\mu}(\vec{r}) , \frac{\tilde{\omega}}{c}
 \vec{\bm{\sigma}}\cdot \vec{I}_A \rho_A (\vec{r}) \rbrace$ \\
  & hyperfine P-even interaction & \\
 \hline
~\\
\multicolumn{3}{c}{Scalar \ptodd term}  \\
$z^{(0,1)}_\mathrm{\cal SP} $ & Scalar \ptodd interaction   & $\mathrm{i}\displaystyle\frac{\fermi}{2\sqrt{2}}2k_\mathrm{{\cal SP},A} Z_{A}[\vec{\bm{\sigma}}\cdot \vec{\bm{p}},\frac{\tilde{\omega}}{c}\rho_A(\vec{r})]$ \\
\end{tabular*}
\end{table*}

Here we briefly outline the fully-relativistic (four-component) operators describing nuclear spin-dependent \podd (\sdpv)
and scalar \ptodd electron-nucleus
interactions and provide the corresponding expression within ZORA in \tref{pvterms} (omitting the response terms).
For detailed discussion on \podd and \ptodd
terms in quasirelativistic approximation see e.g. \cite{berger:2005, nahrwold:09} and \cite{Isaev:2012, Isaev:2013}.
We use atomic units unless other is stated explicitly.

The nuclear spin-dependent \podd and scalar \ptodd operators in four-component
formalism can be written as:
\begin{eqnarray}
 H_\mathrm{\cal SD-PV}=\frac{\fermi}{\sqrt{2}}\sum_{A,i} k_{{\cal A},A}{\vec{\bm{\alpha}}}(i)\cdot \vec{\bf I}_A \rho_A (\vec{r_i}), \\
 H_\mathrm{\cal SP}=\mathrm{i}\frac{\fermi}{\sqrt{2}}\sum_{A,i} k_{\mathrm{\cal SP},A} Z_{A} \bm{\gamma}_0 (i) \bm{\gamma}_5 (i) \rho_A(\vec{r_i}),
 \label{4c-oper}
\end{eqnarray}
where $\fermi=2.22254\times 10^{-14}$a.u. is Fermi's constant of the weak interaction,
$k_{{\cal A},A}$ is an effective parameter describing
\sdpv interactions for nucleus $A$ (caused both by the nuclear anapole moment
\cite{zeldovich:1957,zeldovich:1958,flambaum:1980} and by weak electron-nucleon
interactions \cite{Novikov:1977}), $k_{\mathrm{\cal SP},A}$ is a dimensionless coupling constant describing scalar
 \ptodd interaction for nucleus $A$,
$\vec{\bf I}_A$ and $\rho_A$ are the spin and
normalized nuclear density distribution
of nucleus $A$, respectively,
$\bm{\gamma}_0$ and $\bm{\gamma}_5$ are the Dirac $\gamma$-matrices and $\vec{\bm{\alpha}}$ is a vector
of Dirac's $\alpha$-matricies. The summation goes over all nuclei $A$ and electrons $i$.
In \tref{pvterms}, $Q_A$ is the weak charge of nucleus $A$,
$Q_A=N_A-(1-4\sin^2\theta_\mathrm{W})Z_A$,
where $N_A$ is the number of neutrons in nucleus $A$, $Z_A$ the nuclear
charge, $\sin^2\theta_W$ the Weinberg parameter, for which we employ the
numerical value $\sin^2\theta_\mathrm{W}=0.2319$, and
$\vec{A}_{\mu}$ is the magnetic vector potential from the point-like
nuclear magnetic moments $\vec{\mu}_A = \gamma_A \vec{I}_A$ with
$\vec{A}_{\mu}(\vec{r})= \sum_A \vec{\mu}_A\times
(\vec{r}-\vec{R_A})/(|\vec{r}-\vec{R_A}|)^3$, $\gamma_A$ being the
gyromagnetic ratio,
$\lbrace x,y \rbrace = xy + yx$ is the anticommutator and $[x,y] = xy - yx$
is the commutator. The ZORA factor $\tilde{\omega}$ is also used, $\tilde{\omega}
=1/\left(2-{\widetilde{V}}/c^{2}\right)$, where $\widetilde{V}$
is the model potential (with additional damping \cite{liu:2002}) proposed
in \cite{wullen:1998}, which alleviates the gauge-dependence
of ZORA.

Our ZORA/GHF values (obtained for equilibrium geometry from {\sc ARECP} RaOH calculations)
for  $|W_\mathrm{a}|$ is 1.4$\times$10$^3$ Hz and for  $W_\mathrm{s}$ is 154$\times$10$^3$ Hz.
These values are rather close to the values of the corresponding parameters for RaF from
\cite{Isaev:2013, kudashov:2014}. This is not surprising, 
taking into the consideration that 
the main contribution to the regarded parameters in both cases comes from the unpaired electron 
located on non-bonding molecular
orbital centered on Ra.

\section*{Conclusion}
We propose RaOH as a polyatomic molecule containing
heavy nucleus, particularly suitable for highly effective cooling with lasers.
Using recently developed method for computation of transition matrix elements we calculate
transition dipole moments for cooling transitions for RaOH and estimate Doppler limit
temperature. For the first time \podd and \ptodd properties are calculated for triatomic molecule
with open electronic shells.
Some other polyatomic molecules and ions with heavy nuclei
may be also proposed for direct cooling with lasers
according to the the scheme suggested by us in previous articles.
\section*{Supplementary material}
See supplementary material for the details on electronic structure calculations (basis sets, methods etc)
and extended sets of molecular parameters.
\section*{Acknowledgements}
We are indebted to Prof. R. Berger for providing us with the code hotFCHT for calculations of Franck-Condon factors
in polyatomic molecules. T. I. is especially grateful to R. Berger for numerous insightful discussions during
joint work on calculations of \podd and \ptodd properties for molecules
with open electronic shells and implementing the routines for calculations of these properties withing tm2c code.
We are thankful for Dr. Yu. V. Lomachuk and Dr. Yu. A. Demidov for technical help.
Financial support by RFBR (grants N 16-02-01064 and N 16-03-00766) and computer time
provided by the Center for
Scientific Computing (CSC) Frankfurt is gratefully acknowledged.


\end{document}